\documentclass[proof]{WileyASNA-v1}

\articletype{Article Type}%

\received{05 april 2021}
\revised{11 april 2021}
\accepted{17 april 2021}

\begin{document}

\title{The Branch-cut Cosmology: A topological canonical quantum approach}

\author[1,2]{Peter O. Hess*}

\author[3,4]{C\'esar A. Zen Vasconcellos}

\author[5]{Jos\'e de Freitas Pacheco}

\author[3]{Dimiter Hadjimichef}

\author[6]{Benno Bodmann}

\authormark{P.O. Hess \textsc{et al}}

\address[1]{Universidad Nacional Aut\'onoma de Mexico (UNAM), M\'exico City, M\'exico}

\address[2]{Frankfurt Institute for Advanced Studies (FIAS), Hessen, Germany}

\address[3]{Instituto de F\'isica, Universidade Federal do Rio Grande do Sul (UFRGS), Porto Alegre, Brazil}

\address[4]{International Center for Relativistic Astrophysics Network (ICRANet), Pescara, Italy}

\address[5]{Observatoire de la C\^ote d'Azur, Nice, France}

\address[6]{Universidade Federal de Santa Maria (UFSM), Santa Maria, Brazil}

\corres{*C\'esar A. Zen Vasconcellos. \email{cesarzen@cesarzen.com}}

\abstract{In this contribution we sketch a branch-cut quantum formulation of the Wheeler-DeWitt equation analytically continued to the complex plane. As a starting point, we base our approach on the Ho\v{r}ava-Lifshitz formulation of gravity, which employs higher spatial-derivative terms of the spacetime curvature for renormalisation reasons.
Following standard procedures, the quantization of the Lagrangian density is achieved by raising the Hamiltonian, the dynamical variable $\ln^{-1}[\beta(t)]$, which represents the
the branch-cut complex scale factor, 
and the conjugate momentum $p_{\rm{\ln} }$ to the category of operators. We arrive at an Schrodinger-type equation with a non-linear potential. Solutions are then obtained and discussed for different potential parameterizations. The results reinforce the conception of a quantum leap between the contraction and expansion phases of the branch-cut universe, in good agreement with the Bekenstein criterion.} 


\maketitle

\section{Introduction}

The consistent reconciliation of general relativity (GR)
and quantum mechanics(QM) presents inherent mathematical and conceptual difficulties. The main problem lies in the fact that relativity and quantum mechanics are fundamentally different theories, generating at different scales formal incompatibilities between two genuinely consistent descriptions of reality. 

On the Planck scale, general relativity, based on the geometric properties of a smooth unified spacetime manifold, and quantum mechanics, a probabilistic formulation where time corresponds to an universal and absolute
parameter, are supposed to merge into an unified theory of quantum gravity. 

General relativity provides meaningless answers when reduced to quantum dimensions. Quantum mechanics, on the other hand, lacks conceptual and formal consistency when expanded to cosmic dimensions. 

Furthermore, general relativity describes continuous and deterministic events, reducing causes and effects to a specific location. In quantum mechanics in turn, events produced by the interactions between elementary particles happen in quantum leaps, whose results are probabilistic rather than deterministic. Moreover, in quantum mechanics laws and rules admit connections prohibited by classical physics.

The formal treatment of time in quantum mechanics also differs from that given in general relativity. In quantum mechanics, due to its absolute and universal character, the treatment given to time differs from the treatment given to the other coordinates that unlike time can be elevated to the category of quantum operators and observables. since in QM the time coordinate plays an essentially different role from the position coordinates: while position is represented by a Hermitian operator, time is represented by a c-number\footnote{The principle of correspondence involving classical and quantum theories states that physical systems obey the laws of quantum mechanics with classical physics corresponding to just an approximation valid for large quantum numbers. The quantization procedure in turn correspond to the transition process between both representation structures. In the canonical quantization procedure, the canonical pair of the phase space coordinates ($q,p$) of classical mechanics becomes the position and linear momentum operators ($\hat{q},\hat{p}$) acting linearly on Hilbert quantum-states, which then obey, as a result of quantization procedures, Poisson square brackets which describe canonical commutation relations between $\hat{q}$ and $\hat{p}$.}.

In general relativity, unlike Newtonian physics and quantum mechanics, where the variable time $t$ represents the `reading of a clock', this reading is not given by $t$ but instead by
\begin{equation}
T = \int_{\gamma} \sqrt{g_{\mu\nu}(\vec{x},t) dx^{\mu} dx^{\nu}} \, , \label{T}
\end{equation}
which defines a line integral depending on the gravitational field, $g_{\mu\nu}(\vec{x},t)$, computed along the clock’s worldline $\gamma$~\citep{Rovelli2015}. The coordinate $t$ in the argument of $g_{\mu\nu}(\vec{x}, t)$ despite representing the evolution parameter of the Lagrangian and Hamiltonian formalism, in general relativity has no direct physical meaning, and according to \citet{Rovelli2015}, ``can be changed freely''.

The description of \citet{Rovelli2015} about the role of the variable $t$ in general relativity is very illuminating. Suppose the readings of two clocks, $T_1$ and $T_2$ as given in Equation~\ref{T}. General relativity then describes the relative evolution of these variables computed along different worldliness, with respect to one another, assumed to be on the same level of equality and do not consider the absolute evolution of any of those variables in time. 

From a mathematical point of view, an adequate representation of this relative evolution would be $T_1(T_2)$ and $T_2(T_1)$. However, for simplicity and broad description what general relativity does is replace the arguments of these representations with an arbitrary parameter, $t$, and use the parametric form of representation
$T_1(t)$ and $T_2(t)$, where the variable $t$ can
be chosen freely. In these examples, $t$ is the same parameter that appears in the definition of the gravitational field in Equation~\ref{T}. So, general relativity describes the relative evolution of variable quantities with
respect to one another instead of the exclusive absolute evolution of any of these variables in time (see \citet{Rovelli2019} for more details).

In the following we propose a topological canonical quantum approach for the branch-cut cosmology~\citep{Zen2020,Zen2021a,Zen2021b, Zen2022} based on the geometrodynamics Wheeler-DeWitt equation~\citep{WDW}, based upon canonical quantization of constrained systems~\citep{Shestakova2018}. The solution of the WDW equation, the wave function of the universe, is thus a functional on the geometries of compact three-manifolds and on the values of the matter fields on these manifolds and describes the quantum states of a spatially closed universe~\citep{Hartle}. As a corollary, our expectations concerning this approach are that the Wheeler DeWitt equation, ``properly understood and properly defined'', may give the ``full dynamics
of quantum gravity''~\cite{Rovelli2015}. 

Despite the wide range of meanings of this statement, in the present work, by proposing a quantum formulation of branch-cut cosmology, in this approach we limit ourselves, for a comparison with formulations based on the standard model, to a formulation with a real potential, however extending the formulation to a conjugate representation of the original representation. We thus fulfill one of the most striking conclusions of branch-cut cosmology.

\section{The Weeler and DeWitt Equation}

On basis of the audacious idea of {\it physics without time} as the direct outcome of conversations with Wheeler~\citet{WDW} introduced a canonical formulation of quantum gravity, known today as the Wheeler DeWitt equation. 

Despite the uncertainties involving the original formulation, the main one related to time suppression (see \citet{Rovelli2015}), the WDW equation is considered a milestone in the development of quantum gravity. At present, the absence of time in this equation is perfectly understandable and consistent with current interpretations of the role of time in general relativity\footnote{Technical deficiencies and misinterpretations of the original version have historically led to a tendency to underestimate the WDW equation as a consistent formulation of quantum gravity, despite supporting several approaches, from quantum geometrodynamics to loop quantum gravity. More recently, however, an opposite trend has emerged, related to the understanding of the fundamental reasons for the intriguing explicit absence of the time variable in the WDW equation~\citep{Rovelli2011,Rovelli2015,Shestakova2018}.}.

The Wheeler-DeWitt formulation for quantum gravity consists in constraining a wave-function which applies to the universe as a whole, the so called {\it wave function of the universe}\footnote{The significance of the wave-function of the universe can be ascribed only to the intrinsic dynamics of the universe~\citep{WDW}, or to the probability amplitude for the universe to have some space geometry, or to be found in some point of the Wheeler super-space~\citep{Shestakova2019}.}, in accordance with the Dirac recipe: 
\begin{equation}
\hat{\cal H} \Psi = 0 \, ,
\end{equation}
i.e., a stationary, timeless equation, instead of a time-dependent quantum mechanics
wave equation as for instance
\begin{equation}
i \frac{\partial}{\partial t} \Psi = \hat{H} \Psi \, . 
\end{equation}
Here, $\hat{H}$ denotes the Hamiltonian operator of a quantum subsystem,  
while $\hat{\cal H}$ in the previous equation represents a quantum operator which describes a general relativity constraint, resulting in a second order hyperbolic  equation of gravity variables\footnote{More precisely, the scale factor $a(t)$, the density $\rho(t)$, the pressure $p(t)$, and the gravitation constant $\Lambda$.}. The WDW equations thus constitutes a Klein-Gordon-type equation, having therefore a `natural' conserved  associated current (${\cal J}$),
\begin{equation}
{\cal J} = \frac{i}{2} \Bigl(\Psi^{\dagger} \, \nabla \cdot \Psi -    \Psi \,  \nabla \cdot \Psi^{\dagger} \Bigr); \, \, \, \mbox{with} \, \, \nabla \cdot {\cal J} = 0 \, . \label{J}  
\end{equation}

The Wheeler-DeWitt equation presents some notable features that are distinguishable from the Schr\"odinger equation. The main feature, as stressed before,  is the absence of the time coordinate such that the `universe wave-function $\Psi$' depends only on the three-dimensional spatial coordinates, giving rise to the well-known `time problem'\footnote{The so-called `time problem' designates the conceptual conflict between quantum mechanics and general relativity, in which the former considers the flow of time to be universal and absolute, while for the latter time is relative. The problem of time also involves the issue related to the flow of time, as in general relativity time seems to flow in a single direction while in quantum mechanics the flow of time is symmetrical.}~\citep{Rovelli2011,Rovelli2015}. 

The main reason for the absence of a specific external coordinate of time lies in the covariance imposition of general relativity, more precisely, the wave
function is only a function of the “3-geometry”, due namely to
the equivalence class of metrics under a diffeomorphism\footnote{In mathematics, a diffeomorphism is an isomorphism of smooth manifolds. It is an invertible function that maps one differentiable manifold to another such that both a given function and its inverse are differentiable.} (for the details see \citet{Rovelli2015}). The second important feature is that the original version of the WDW wave equation is real instead of complex. This characteristic is due to the intrinsic union between the time variable and the imaginary unit $i$ in the definition of space-time\footnote{It is important to remember that the presence of the imaginary unit i in the Schr\"oringer equation is fundamental to guarantee the conservation of probability.}. 
Evidently, in what follows, in view of the complexification of the FLRW metric in the branch-cut cosmology, the corresponding WDW-type equation presents different conceptual components with respect to both the meaning of the wave function of the universe, as well as the understanding of the corresponding continuity equation, among others. 

 Based on this understanding, in the following we sketch a branch-cut quantum formulation of the WDW equation analytically continued to the complex plane, characterized by only one dynamical variable, the branch-cut complex scale factor $\ln^{-1}[\beta(t)]$ (see \citep{Zen2020,Zen2021a,Zen2021b}).

 \section{Ho\v{r}ava-Lifshitz action}
 
 The Ho\v{r}ava-Lifshitz formulation of gravity~\citep{Horava,Cordero2019,Compean,Garattini2016,Vakili,Chojnacki} is an alternative theory to general relativity which
employs higher spatial-derivative terms of the spacetime curvature which are added to the Einstein-
Hilbert action with the aim of obtaining a renormalisable theory. 
The Ho\v{r}ava-Lifshitz (HL) gravity model preserves high and low energy scales and proposes the ultraviolet completion and a power-countable renormalization of gravity at the UV fixed point, additionally recovering the standard theory at the fixed infrared point, as well as Lorentz symmetry.
Hořava–Lifshitz topological gravity addresses, in its original formulation, quantum mechanics as the fundamental theoretical framework so that space and time are not equivalent and therefore anisotropic\footnote{Renormalization of the theory at high energies despite causing the breaking of Lorentz symmetry, whose validity domain corresponds to low energies, is achieved by the imposition of an anisotropic scale between space and time, leading in turn to the ADM 3+ 1 spliting~\citep{ADM}, and a foliation that preserves the diffeomorphism, parameterized by a global time~\citep{Horava}.}.
 
 In the following, as a simplification, instead of adopting a superspace formulation\footnote{The construct {\it superspace} corresponds to the space that comprises all Riemannian three-metrics configurations, $h_{ij}(\vec{x})$, and matter field configurations, $\Phi(\vec{x})$, on a spatial hypersurface, $\Sigma$
$Riem(\Sigma) = \{h_{ij}(\vec{x}, \Phi(\vec{x}) | \vec{x} \in \Sigma  \}$. Given a diffeomorphism relating a set of configurations, they are considered equivalent since they have the same intrinsic geometry and therefore this space can be divided into equivalence classes.
We then identify a superspace as $Sup(\Sigma) = Riem(\Sigma)/Diif_0(\Sigma)$, where the zero subscript indicates diffeomorphisms connected only to the identy and the metric of the infinite dimensional superspace corresponds to the Wheeler-DeWitt metric.} we assume as a starting point a homogeneous and isotropic multiverse restricted to
a mini-superspace\footnote{The Wheeler-DeWitt equation does not actually represent a single equation, but an infinite number of equations, one for each point x of a spatial hypersurface $\Sigma$, making its resolution extremely difficult. A simplification corresponds to considering a finite number of degrees of freedom of the metric and matter fields realization in a mini-superspace. This simplification limits the obtainment of an exact solution for the Wheeler-DeWitt equation, only approximate, but allows the knowledge of its fundamental properties~\citep{Flinn}.}  (see for instance~\citep{Kim}), corresponding to a 4-dimensional space-time branch-cut universe manifold, $\cal{M}$, decomposed into $3+1$ dimensions\footnote{Such a decomposition is possible in general if the manifold $\cal{M}$ is globally hyperbolic~\citep{Wiltshire}.}.

The branch-cut cosmology is characterized by only one dynamical variable, the branch-cut complex scale factor $\ln^{-1}[\beta(t)]$. The branch-cut manifold $\cal{M}$ is foliated into hypersurfaces, $\Sigma_t$, constricted to Riemann sheets labeled by a complex parameter of time, $t$, with the 4-dimensional  analytically continued branch-cut line element given as~\citep{Zen2021a, Zen2021b}
\begin{equation}
ds_{[\rm{ac}]}^2     =    -  \sigma^2   N^2(t)  c^2dt^2   +    \sigma^2  \ln^{-2}[\beta(t)] d\Omega^2(r,\theta,\phi) \, ; \label{FLRWac2}  
\end{equation}
the metric of the unit 3-sphere is
\begin{equation}
d\Omega^2(r,\theta,\phi) \equiv
 \Biggl[ 
\frac{dr^2}{\bigl(1 -  kr^2(t) \bigr)}
  +   r^2(t) \Bigl(d \theta^2   +   sin^2 \theta d\phi^2  \Bigr)  \Biggr] \, . \label{3metric}
\end{equation}
In this expression, the function $\beta(t)$ characterizes the range of $\ln^{-1}[\beta(t)]$ associated to the cuts in the branch-cut. The branch-cut complex scale factor $\ln^{-1}[\beta(t)]$ denotes solutions of Einstein's equations for a FLRW-type metric extended to the complex plane, represented as the reciprocal of a complex multi-valued function, the natural complex logarithm function $\ln[\beta(t)]$, not its inverse. 
Moreover, in these expressions, $r$ and $t$ represent respectively real-space and complex-time parameters and $k$ encodes the spatial curvature of the multi-composed universe, $k = -1, 0, 1$ for, respectively, negatively curved, flat or positively curved spatial hyper-surfaces analytically continued to the complex plane. 

In expression (\ref{FLRWac2}), $N(t)$ is an arbitrary lapse function  and 
$\sigma^2 = 2/3\pi$ is just a normalisation factor. 
The lapse
function $N(t)$ is not a dynamical quantity, but a pure gauge variable. Gauge invariance of the action in general relativity yields a
Hamiltonian constraint which requires a gauge fixing condition on the lapse $N(t)$ (see~\citet{Feinberg}.
The lapse function may be fully determined by the difference between the elapsed time-parameter $t$, and the proper time-parameter $\tau = N(t) dt$ on curves normal to a given hypersurface  $\Sigma_t$~\citep{Flinn,Wiltshire}.  A canonical analysis shows that the lapse
function $N(t)$ and the ADM~\citep{ADM} shift vector $N^i(t)$ play the role of Lagrange multipliers~\citep{Kim}. In the special case in which $N^i = 0$, the spatial coordinates are said to be “comoving”. The 
shift vector $N_i$ describes how a given hyper-surface $\Sigma_t$ differs from a neighbouring hyper-surface $\Sigma_{t +dt}$. In this case 
the time-parameter component of the line element would be shifted to $N^2(t) \to N^i N_i$. 

The extrinsic curvature\footnote{Unlike intrinsic curvature, extrinsic curvature of a given sub-manifold of a manifold is a curvature property which depends on its particular embedding.
In the particular case of an immersed sub-manifold of a Euclidean space $\Re^N$ this means a differentiable manifold ${\cal M}$ together with an immersion $X : {\cal M} \to \Re^N$.
} is a symmetric, purely spatial mathematical construct that measures the geometrical changes and deformation rates of the normal to a hyper-surface as it is transported from one point to another. Extrinsic curvature corresponds to a manifold (actually a sub-manifold) which depends on its particular embedding. 
The extrinsic curvature tensor $K_{ij}$ can be written in terms of the three-metric in the following form~\citep{Bertolami2011}:
\begin{equation}
K_{ij} =\frac{1}{2\sigma N} \Biggl(- \frac{\partial g_{ij}}{\partial t} + \nabla_i N_j +  \nabla_j N_i  \Biggr)   \, ,   
\end{equation}
where $N^i$ $\nabla_i$ denotes the 3-dimensional covariant derivative, and
$g_{ij} =  \ln^{-1}[\beta(t)]  diag \bigl(\frac{1}{1 - r^2}, r^2, r^2sin^2 \theta \bigr)$. 

Since $N_i = 0$ for RW-like spaces, corresponding to co-moving coordinates, the curvature tensor takes the form
\begin{equation}
    K_{ij} = - \frac{1}{2\sigma N} \frac{\frac{d}{dt} \ln^{-1}[\beta(t)]}{\ln^{-1}[\beta(t)]}g_{ij} \, .
\end{equation}    
Taking the trace of the curvature tensor we get
  \begin{equation}
K = K^{ij} g_{ij} =  - \frac{3}{2\sigma N} \frac{\frac{d}{dt} \ln^{-1}[\beta(t)]}{\ln^{-1}[\beta(t)]}.
\end{equation}
The next step is to write the Ho\v{r}ava-Lifshitz formulation of the Einstein–Hilbert action as a functional formulation of the spacetime metric $g_{\mu \nu}$. The dynamics of the system corresponds to a succession of three-dimensional space-like hypersurfaces imbedded in a four-dimensional space-time. 

The Ho\v{r}ava-Lifshitz action for an isotropic and homogeneous mini-superspace is defined as the Lagrangian density:

\begin{eqnarray}
& \!\!\!\!\!\!\!\!{\cal S}_{HL}  \!\! =   \frac{M_P}{2} \! \! \int \!\! d^3x dt N \sqrt{-g} \Bigl\{K_{ij}K^{ij} \! - \! \lambda K^2 - \eta_0 M^2_p  -  \eta_1 R & \nonumber \\
& \!\!\!\!\!\! -  \, \eta_ 2 M^{-2}_P \! R^2  -  \eta_3 M^{-2} \! R_{ij} R^{ij}  -   \eta_4 M^{-4}_P \! R^3  -  \eta_5 M^{-4} \! R(R^i_j R^j_i) & 
 \nonumber \\ & \!\!\!\! - \,  \eta_6 M^{-4} R^i_j R^j_k R^k_i  -  \eta_7 M^{-4}_P R \nabla^2 \! R   -  \eta_8 M^{-4}_P \nabla_i R_{jk} \nabla^i R^{jk}  \Bigr\},  & \nonumber \\
 \label{SHL} 
\end{eqnarray}
where $\eta_i$ denote coupling constants, $M_P$ is the Planck mass, and  $\nabla_i$ represent covariant derivatives. Next, we assume values for some of the coupling constants conventionally adopted in general relativity: $\eta_1 = -1$ (as a result of rescaling the time parameter), and $\eta_0 M^2_p = 2 \Lambda$ where $\Lambda$ denotes the cosmological constant. 
In expression (\ref{SHL}) the branch-cut Ricci components of the 3-metrics are obtained as the foliation in a surface of maximum symmetry\footnote{We do not perform any alteration of the $\lambda$ coupling constant at this point, but it is important to stress that in general relativity  $\lambda = 1$ corresponds to the full diffeomorphism invariance. Moreover, since the Ho\v{r}ava-Lifshitz action proposal introduces only an anisotropy between space and time, it does not alter the homogeneity of the conventional FLRW metric.}, 

\begin{equation}
    R_{ij} = \frac{2}{\sigma^2 \bigl( \ln^{-1}[\beta(t)]\bigr)^2} g_{ij}\, , \quad  \mbox{with} \quad R = \frac{6}{\sigma^2 \bigl( \ln^{-1}[\beta(t)]\bigr)^2} ,
\end{equation}
with $R$ denoting the scalar curvature. 

This parametrization lead to the following expression for the action \ref{SHL}

\begin{eqnarray}
& \!\!\!\!\! {\cal S}_{HL}  \! = \!    \frac{M_P}{2} \!\! \int  \! d^3x dt N \sqrt{-g} \, \Bigl\{ K_{ij}K^{ij}  -  \lambda K^2  + R  -  2 \Lambda   & \nonumber \\
 & \!\!\!\! - \, \eta_ 2 M^{-2}_PR^2  -  \eta_3 M^{-2} \! R_{ij} R^{ij} \! -   \eta_4 M^{-4}_P \! R^3 \! -  \eta_5 M^{-4} \! R(R^i_j R^j_i) &
 \nonumber \\ & \!\!\!\!\! -  \, \eta_6 M^{-4} R^i_j R^j_k R^k_i \! -  \eta_7 M^{-4}_P R \nabla^2 \! R  \! -  \eta_8 M^{-4}_P \nabla_i R_{jk} \nabla^i R^{jk} \! \Bigr\}. &  \nonumber \\ \label{SHL2}
\end{eqnarray}

The subsequent steps of the formal approach follow a standard procedure: firstly we combine equation (\ref{SHL2}) with the metric expression (\ref{3metric}), then we perform the integral\footnote{In Ho\v{r}ava-Lifshitz cosmology, in the case where the lapse function, $N(\vec{r},t)$, is only a function of time, as generally assumed, it implies that the classical Hamiltonian of general relativity is non-local, and must be integrated over the spatial coordinates~\citep{Mukohyama,Saridakis}. According to~\citet{Mukohyama2009}, this condition generates an additional term in the Hamiltonian that mimics the presence of ``dark dust'' in Friedman's equations. However, since the Robertson-Walker metric is homogeneous, this spatial integral simply corresponds to the spatial volume of space and, therefore, the presence of “dark dust” must disappear~\citep{Maeda,Bertolami2011}.} $\int d^3 x \sqrt{g} = 2 \pi^2 \bigl(\ln^{-1}[\beta(t)]\bigr)^3$, and finally we recombine the Ho\v{r}ava-Lifshitz action coupling constants\footnote{Here, $\eta_c > 0$, stands for the curvature coupling constant~\citep{Bertolami2011}; $\eta_r$ corresponds to the coupling constant for the radiation contribution and 
$\eta_s$ stands for the ``stiff'' matter contribution (which corresponds to the $\rho = p$ - equation of state);
$\eta_r$ and $\eta_s$ can be either positive
or negative since their signal does not alter the stability of the 
Ho\v{r}ava-Lifshitz gravity~\citep{Bertolami2011}. }. 
This procedure allows to obtain the following expression for the Lagrangian density:

\begin{eqnarray}
 {\cal S}_{HL} & \! = \! & \frac{1}{2} \!\! \int \!\! dt \biggl( \frac{N}{\ln^{-1}[\beta(t)]} \biggr)   \!\! \Biggl[\! - \biggl( \frac{\ln^{-1}[\beta(t)]}{N}\biggr)^{\!2} \! \biggl(\frac{d}{dt} \ln^{-1}[\beta(t)]\biggr)^{\! 2}  
\nonumber \\
 &\!+\!& \eta_c \biggl(\! \ln^{-1}[\beta(t)] \! \biggr)^2 \!\!\! - \!  \eta_{\Lambda} \biggl(\! \ln^{-1}[\beta(t)]\! \biggr)^4 \!\!\! - \! \eta_r  \! - \! \frac{\eta_s}{\bigl(\ln^{-1}[\beta(t)]\bigr)^2} \! \Biggr].
 \nonumber \\
\end{eqnarray}

From this expression we can simply read off the Lagrangian
density of the multiverse combining a non-interaction part part, ${\cal L}_0$, and an interacting part, ${\cal L}_{\rm int}$:
\begin{equation}
\!\!\! {\cal L}  =   {\cal L}_0  + {\cal L}_{\rm int}  \, ,
\end{equation}
with
\begin{eqnarray}
 {\cal L}_0 \! = \! - \frac{1}{2} \biggl(\! \frac{N}{\ln^{-1}[\beta(t)]} \! \biggr) \!  \Biggl[\! \biggl(\! \frac{\ln^{-1}[\beta(t)]}{N} \! \biggr)^{\!2} \! \biggl(\frac{d}{dt} \ln^{-1}[\beta(t)] \! \biggr)^{2}  \biggr] , 
\end{eqnarray}
and
\begin{eqnarray}
\!\!\!\! {\cal L}_{\rm int}  \! =  \! \frac{1}{2} \biggl(\! \frac{N}{\ln^{-1}[\beta(t)]} \! \biggr) \! \Biggl[
  \eta_C  \! \biggl(\ln^{-1}[\beta(t)] \! \biggr)^2 \!\! \!\! & \! - \!  & \! \eta_{\Lambda} \! \biggl( \ln^{-1}[\beta(t)] \biggr)^4 \nonumber \\
  - \eta_r  & \! - \! & \frac{\eta_s}{\bigl(\ln^{-1}[\beta(t)]\bigr)^2} \! \Biggr].
\end{eqnarray}

In the following, on basis of this Lagrangian formulation, we proceed with the quantisation of the system.

\section{Topological Canonical Quantisation}\label{QF}

The branch-cut scale factor $\ln^{-1}[\beta (t)]$ as the only dynamical variable of the model may be raised, as we stressed before, in a quantum approach, at the level of a quantum operator. This new status gives the scale factor $\ln^{-1}[\beta (t)] $ an additional role in describing the evolutionary process of the universe, that of representing the formal confluence between the classic and quantum topological description in the branch-cut cosmology. We call this procedure a {\it spacetime topological quantisation}\footnote{Although such a denomination is not the main focus of this work, it is still important to preliminary present some justifications for the adoption of this nomenclature. We assume that the spacetime that supports the branch-cut cosmology represents a topological space, thus consisting of the combination of a manifold and a specific topology. This is because, as we know, every metric space is a topological space where a given topology is induced by the metric. Furthermore, going further and considering the quantum extension of branch-cut cosmology and a discrete topology, we assume that local subsets are equally metrisable and that their topological properties are induced by the main metric.}, which characterises a new perspective in the quantisation of spacetime. 
Our formulation describes in short the relative evolution of the variable $\ln^{-1}[\beta(t)]$ over worldlines $\gamma_{\rm{\ln}} $ associated with hypersurfaces $\Sigma_{\rm{\ln}}$ analytically continued to the complex plane. 

The conjugate momentum $p_{\rm{\ln}}$ of the dynamical variable $\ln^{-1}[\beta(t)]$ is 
\begin{equation}
p_{\rm{\ln}} = \frac{\partial {\cal L}_0}{\partial \bigl(\frac{d}{dt} \ln^{-1}[\beta(t)]\bigr)} =  - \frac{\ln^{-1}[\beta(t)]}{N}\frac{d}{dt} \ln^{-1}[\beta(t)] \,.
\end{equation}
Therefore applying a standard formalism in quantum mechanics,  the branch-cut Hamiltonian becomes
\begin{eqnarray}
 {\cal H}  & = & p_{\rm{\ln}} \frac{d}{dt} \ln^{-1}[\beta(t)]  - {\cal L}  \nonumber \\
& = &  \frac{1}{2} \frac{N}{\ln^{-1}[\beta(t)]} \Biggl(- p^2_{\ln}  + \eta_c \biggl( \ln^{-1}[\beta(t)]  \biggr)^2  \nonumber \\   & - &    \eta_{\Lambda} \biggl( \ln^{-1}[\beta(t)] \biggr)^4
 - \eta_r  \! - \! \frac{\eta_s}{\bigl(\ln^{-1}[\beta(t)]\bigr)^2}  \Biggr) \, .  \label{H}
\end{eqnarray} 

The quantisation of the Lagrangian density is achieved by raising the Hamiltonian, the dynamical variable $\ln^{-1}[\beta(t)]$ and the conjugate momentum $p_{\rm{\ln}}$ to the category of operators   
in the form
\begin{equation}
 \ln^{-1}[\beta(t)] \rightarrow \hat{\ln}^{-1}[\beta(t)] \, ,
\end{equation}
and
\begin{equation}
p_{\rm{\ln}} \rightarrow \widehat{p}_{\rm{\ln}} = - i\hbar \frac{\partial}{\partial \ln^{-1}[\beta(t)]} ; 
\end{equation}
(for simplicity, in the following we skip using the hat symbol in the operators $\hat{\rm{p}}$ and  $\hat{\ln}$).
Ambiguities in ordering of the operators may be overcome by the introduction of an ordering-factor
$\alpha$ in the form
\begin{equation}
p^2 = - \frac{1}{\ln^{-\alpha}} \frac{\partial}{\partial \ln^{-1}[\beta(t)]} \Biggl( \bigl( \ln^{-1}[\beta(t)] \bigr)^{\alpha}\frac{\partial}{\partial \ln^{-1}[\beta(t)]} \Biggr) \, , \label{p2}
\end{equation}
with $\alpha$ usually chosen in general as $\alpha = [0,1]$;  $\alpha = 0$ corresponds to the semi-classical value; intermediate values have no meaning.  

Combining (\ref{H}) and (\ref{p2}) and changing variable\footnote{It is important to mention that the variable $u$ as well as $\ln^{-1}[\beta(t)]$ are complex.} ($u \equiv \ln^{-1}[\beta(t)]$ with $du \equiv d\ln^{-1}[\beta(t)]$), we obtain the following expression for the WDW differential equation:
\begin{equation}
 \Biggl(-\frac{1}{u^{\alpha}}\frac{d}{d u} \Bigl(u^{\alpha}\frac{d}{d u} \Bigr) +  \eta_c u^2  - \eta_{\Lambda} u^4 
 - \eta_r   -  \frac{\eta_s}{u^2}  \Biggr) \Psi(u) = 0 \, ,
 \end{equation}
 where $\Psi(u)$ is the wave function of the Universe. 
From this expression, using
$\alpha~=~0$, we get
\begin{equation}
 \Biggl(- \frac{d^2}{d u^2} +  \eta_c u^2  - \eta_{\Lambda} u^4 
 - \eta_r   -  \frac{\eta_s}{u^2}  \Biggr) \Psi(u) = 0 \, . \label{WDWS}
 \end{equation}
This equation resembles a Schr\"odinger-type equation under the action of a  WDW quantum real potential\footnote{As previously mentioned, in this work, for comparison purposes, we limit ourselves to formulating solutions for the real part of the potential, despite the fact that the variable $u$ is complex. We will discuss the presence of an imaginary component in the potential term in the future.} $V(u)$ acting on the wave-function of the branch-cut universe:
\begin{equation}
     \Biggl(- \frac{d^2}{d u^2} + V(u)  \Biggr) \Psi(u) = 0
\end{equation}
with
\begin{equation}
V(u) = - \eta_{\Lambda} u^4  +   \eta_c u^2
 -  \frac{\eta_s}{u^2} -\eta_r \, , 
\end{equation}
or, in terms of the complex branch-cut scale factor:
\begin{eqnarray}
V(\ln^{-1}[\beta (t)]) =  & - & \eta_{\Lambda} \bigl( \ln^{-1}[\beta (t)]\bigr)^4  +   \eta_c \bigl( \ln^{-1}[\beta (t)]\bigl)^2
\nonumber \\
& - & \frac{\eta_s}{\bigl( \ln^{-1}[\beta (t)]\bigr)^2} -\eta_r \, . \label{Pot}
\end{eqnarray}

\begin{figure}[htpb]
\centering
\includegraphics[width=50mm,height=45mm]{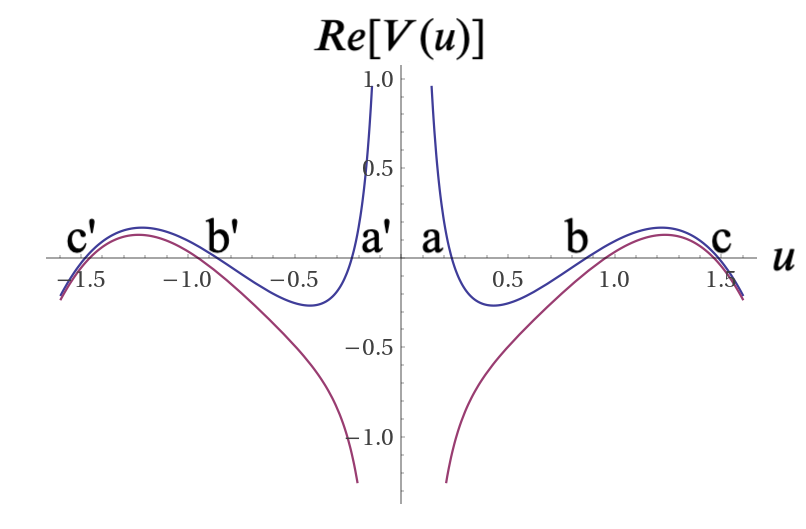}
\caption{Plot of $Re[V(u)]$, 
from equation~\ref{Pot}. In the upper figures the values of the constants are: $\eta_{\Lambda} = 1/3$, $\eta_c = 1$, $\eta_s = -0.03$, and $\eta_r = 0.6$.
In the lower figures the values of the constants are: $\eta_{\Lambda} = 1/3$, $\eta_c = 1$, $\eta_s = +0.03$, and $\eta_r = 0.6$.
Values of parameters  taken from~\citep{Cordero2019}.
} \label{V}
\end{figure}
\begin{figure}[htb]
\centering
\includegraphics[width=50mm,height=45mm]{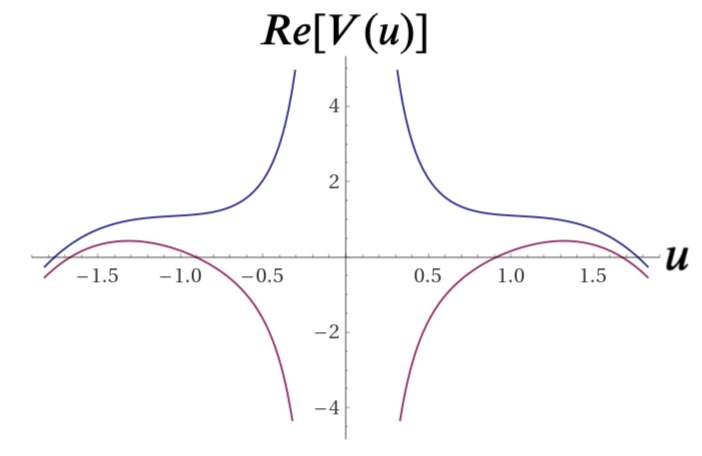}
\caption{Plot of $Re[V(u)]$, 
from equation~\ref{Pot}. In the upper figures the values of the constants are: $\eta_{\Lambda} = 1/3$, $\eta_c = 1$, $\eta_s = -0.468$, and $\eta_r = 0.024$.
In the lower figures the values of the constants are: $\eta_{\Lambda} = 1/3$, $\eta_c = 1$, $\eta_s = +0.468$, and $\eta_r = 0.024$.
Values of parameters  taken from~\citep{Cordero2019}.
} \label{V2}
\end{figure}
\begin{figure}[htpb]
\centering
\includegraphics[width=50mm,height=45mm]{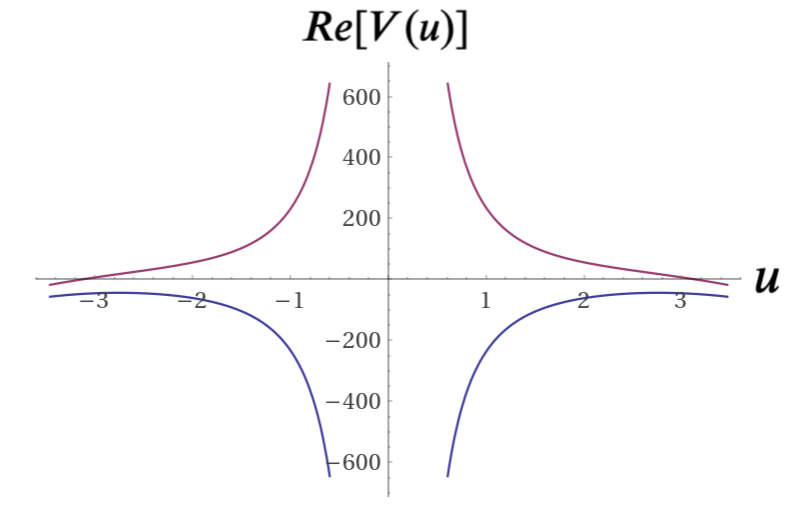}
\caption{Plot of $Re[V(u)]$, 
from equation~\ref{Pot}. In the upper figures the values of the constants are: $\eta_{\Lambda} = 1/3$, $\eta_c = 1$, $\eta_s = -234.0$, and $\eta_r = 0$.
In the lower figures the values of the constants are: $\eta_{\Lambda} = 1/3$, $\eta_c = 1$, $\eta_s = +234.0$, and $\eta_r = 0$.
Values of parameters  taken from~\citep{Cordero2019}.
} \label{V3}
\end{figure}
\begin{figure}[htpb]
\centering
\includegraphics[width=60mm,height=50mm]{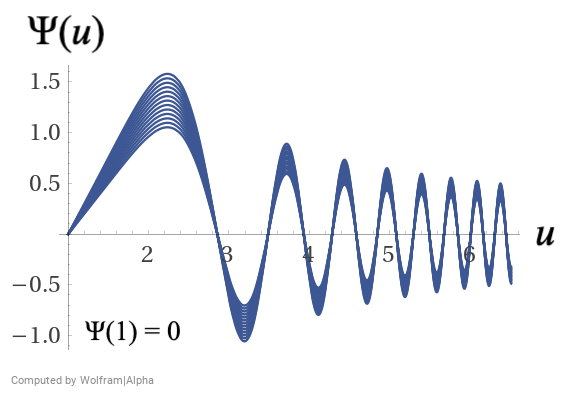}
\caption{Plot of a family of solutions of Equation \ref{WDWS} with initial condition $\Psi(1)=0$.
The values of the constants are: $\eta_{\Lambda} = 1/3$, $\eta_c = 1$, $\eta_s = +0.03$, and $\eta_r = 0.6$. Values of parameters  taken from~\citep{Cordero2019}.} \label{Sol1}
\end{figure} 
\begin{figure}[htpb]
\centering
\includegraphics[width=60mm,height=50mm]{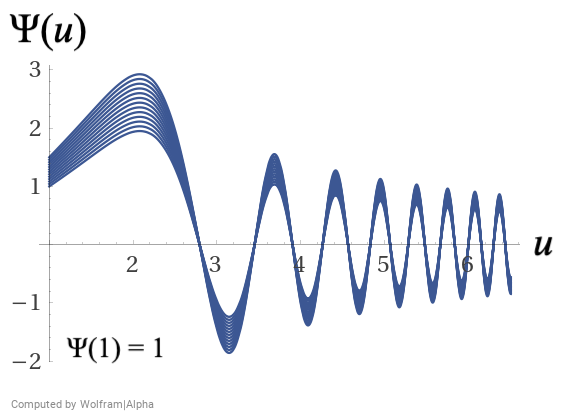}
\caption{Plot of a family of solutions of Equation \ref{WDWS} with initial condition $\Psi(1)=1$.
The values of the constants are: $\eta_{\Lambda} = 1/3$, $\eta_c = 1$, $\eta_s = +0.03$, and $\eta_r = 0.6$. Values of parameters  taken from~\citep{Cordero2019}.} \label{Sol2}
\end{figure} 
In Figure (\ref{V}) we sketch the behavior of the real part of the potential given by the equation \ref{Pot} in two extreme cases, corresponding to $\eta_s > 0$ and $\eta_s <0$. The values of the parameters  were taken from~\citep{Cordero2019} and correspond to typical values of approaches to formulating the Wheeler-DeWiit equation based on general relativity. The treated cases present a kind of potential barrier whose overcoming can occur in the form of quantum tunneling. In the case of the upper figures, both universes, in the contraction and expansion phases, in the case of no quantum tunneling, oscillate around two points of intersection with the horizontal axis, a' and b' and a and b, respectively. Morover, the upper figures exhibit infinite potential barriers that prevent the transition between the two phases, except in the case of a quantum leap. The lower figures do not present this infinite potential barrier but singularities. In the upper figures, singularities are suppressed.

\begin{figure*}[htpb]
\centering
\hspace{-0.1in} \includegraphics[width=43.5mm,height=35mm]{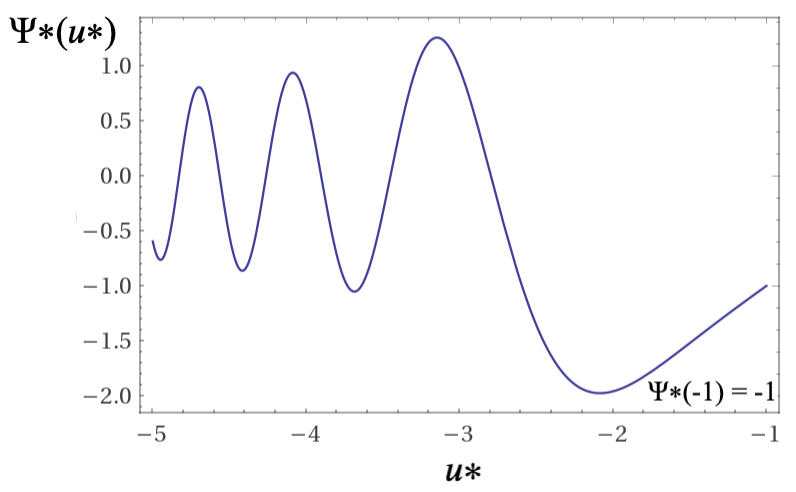} 
\includegraphics[width=43.5mm,height=35mm]{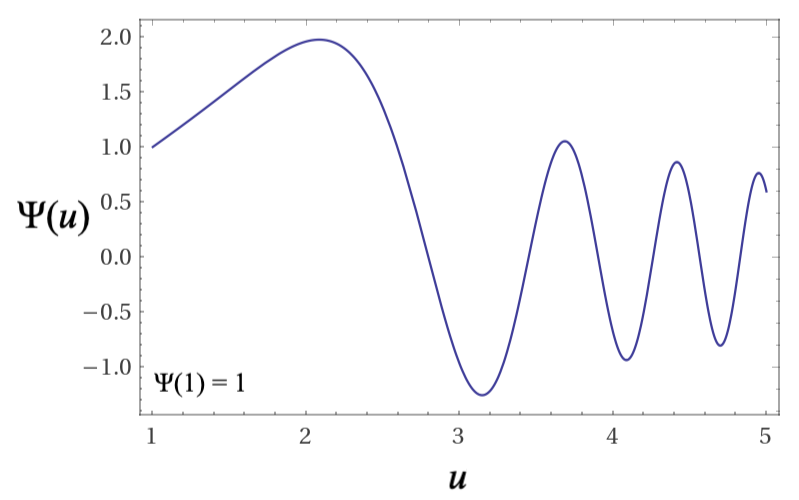} \hspace{0.1in}
\includegraphics[width=43.5mm,height=35mm]{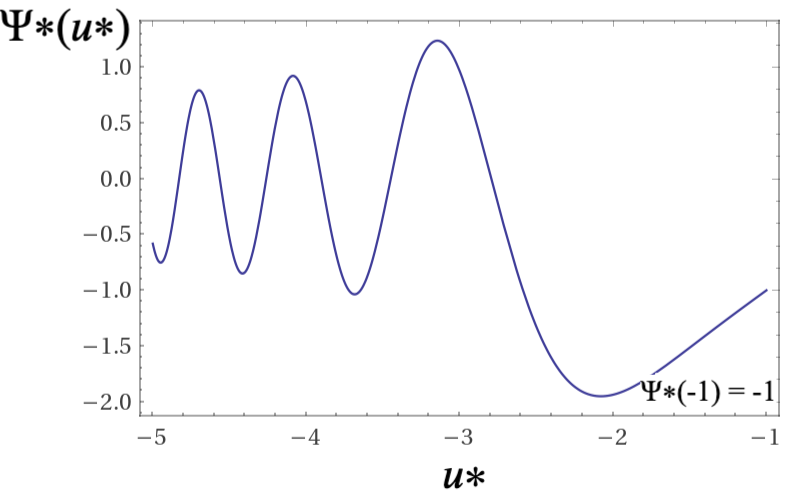} 
\includegraphics[width=43.5mm,height=35mm]{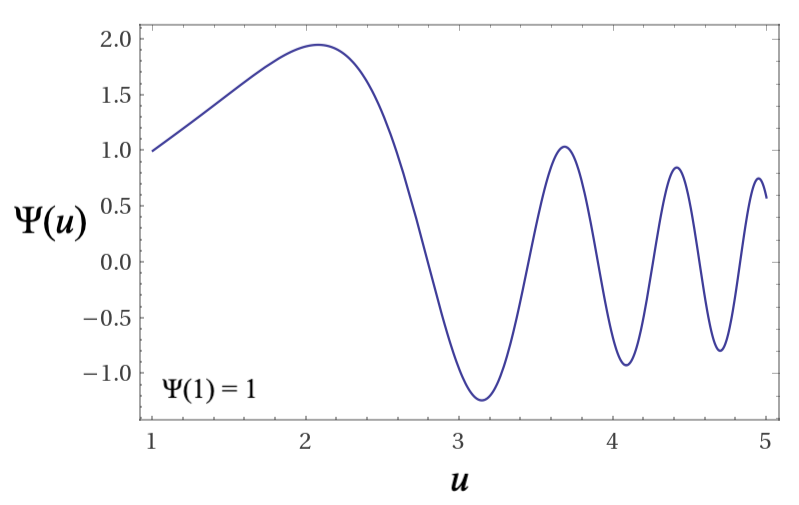}
\caption{In the figures, plot of typical solutions of Equations \ref{WDWS} and \ref{WDWScc} with boundary condition $\Psi(-1)=-1$ on the negative sector and  $\Psi(1) = 1$ on the positive sector.
The values of the constants on the figures on the left are: $\eta_{\Lambda} = 1/3$, $\eta_c = 1$, $\eta_s = -0.03$, and $\eta_r = 0.6$. In the figures on the right, the values of the constants are: $\eta_{\Lambda} = 1/3$, $\eta_c = 1$, $\eta_s = 0.03$, and $\eta_r = 0.6$. Values of parameters  taken from~\citep{Cordero2019}.} \label{LeftRight}
\end{figure*} 

\begin{figure*}[htpb]
\centering
\hspace{-0.1in} \includegraphics[width=43.5mm,height=35mm]{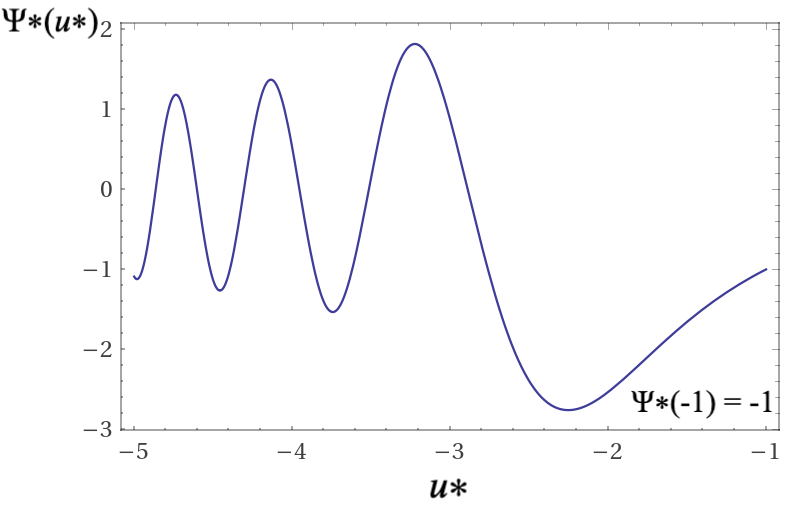} 
\includegraphics[width=43.5mm,height=35mm]{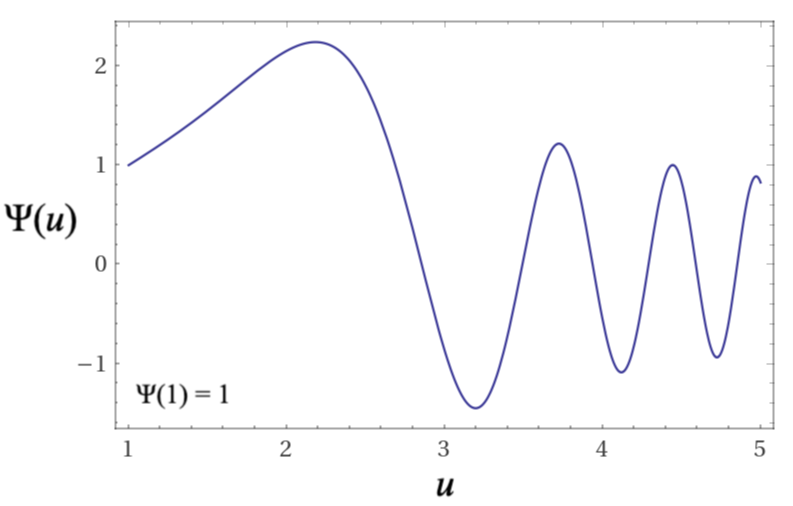} \hspace{0.1in}
\includegraphics[width=43.5mm,height=35mm]{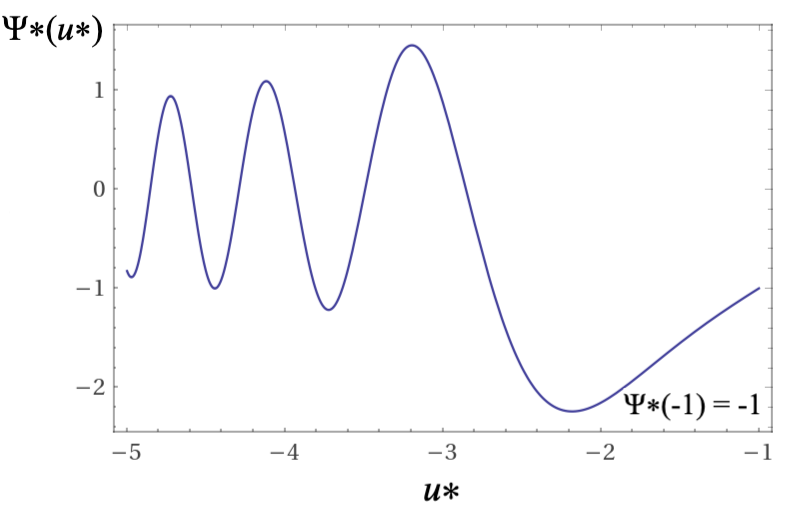} 
\includegraphics[width=43.5mm,height=35mm]{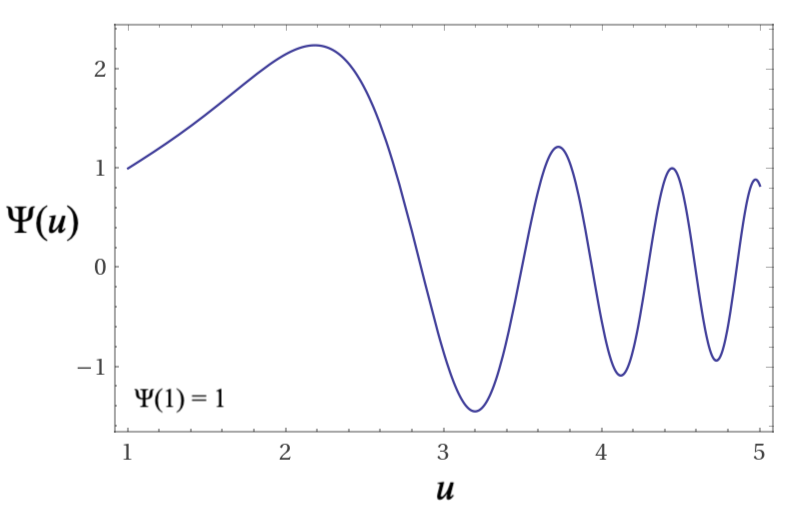}
\caption{In the figures, plot of typical solutions of Equations \ref{WDWS} and \ref{WDWScc} with boundary condition $\Psi(-1)=-1$ on the negative sector and  $\Psi(1) = 1$ on the positive sector.
The values of the constants on the figures on the left are: $\eta_{\Lambda} = 1/3$, $\eta_c = 1$, $\eta_s = -0.468$, and $\eta_r = 0.024$. In the figures on the right, the values of the constants are: $\eta_{\Lambda} = 1/3$, $\eta_c = 1$, $\eta_s = +0.468$, and $\eta_r = 0.024$. Values of parameters  taken from~\citep{Cordero2019}.} \label{LeftRight2}
\end{figure*} 

\begin{figure*}[htb]
\centering
\hspace{-0.1in} \includegraphics[width=43.5mm,height=35mm]{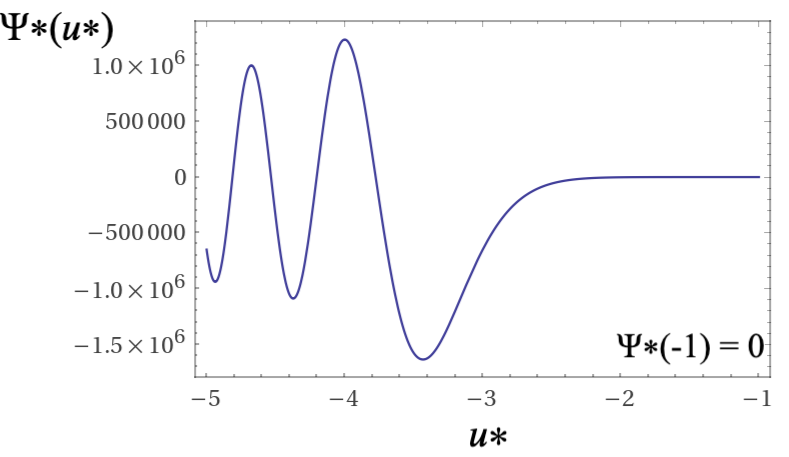} 
\includegraphics[width=43.5mm,height=35mm]{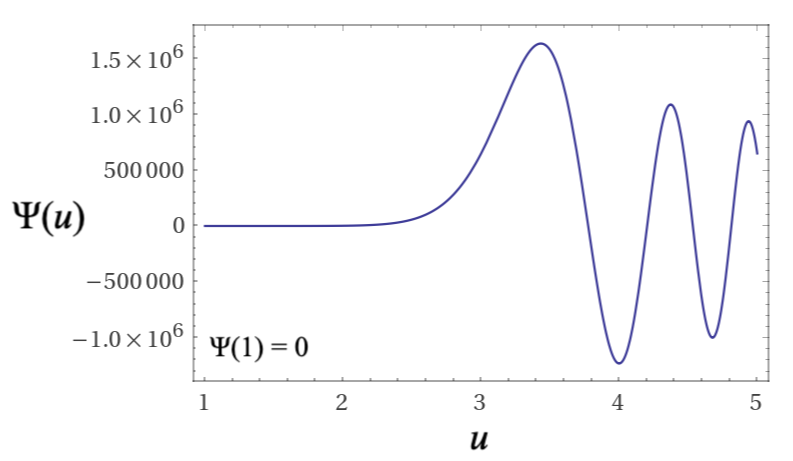} \hspace{0.1in}
\includegraphics[width=43.5mm,height=35mm]{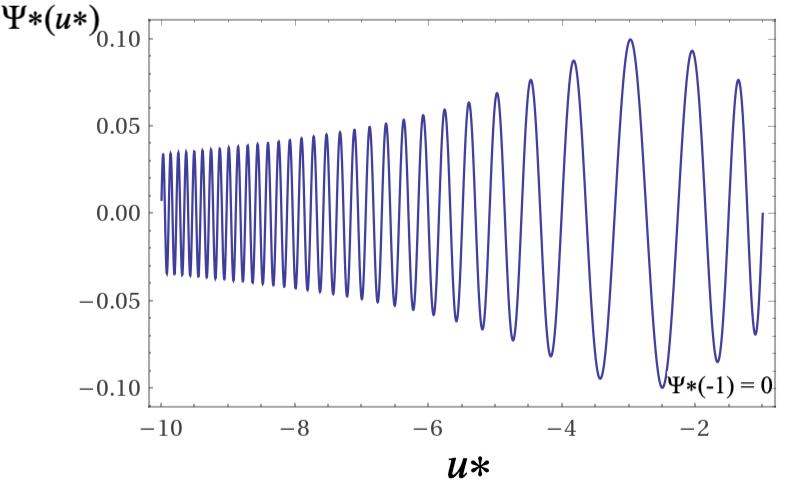} 
\includegraphics[width=43.5mm,height=35mm]{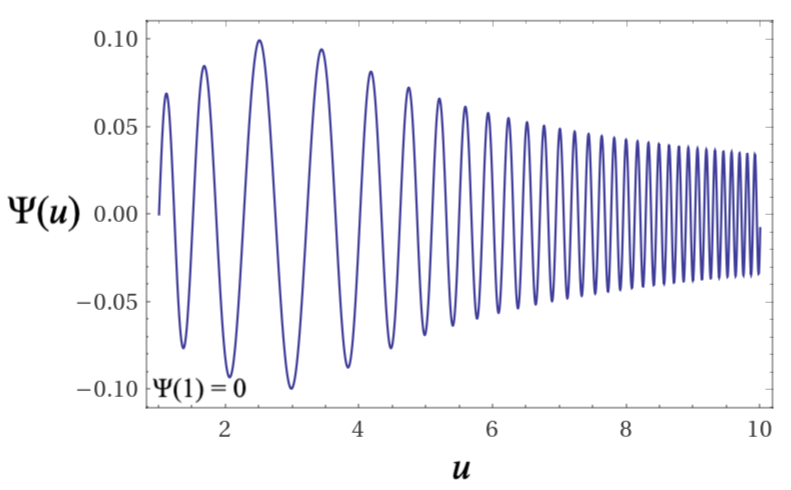}
\caption{
In the figures, plot of typical solutions of Equations \ref{WDWS} and \ref{WDWScc} with boundary condition $\Psi(-1)=0$ on the negative sector and  $\Psi(1) = 0$ on the positive sector.
The values of the constants on the figures on the left are:
$\eta_{\Lambda} = 1/3$, $\eta_c = 1$, $\eta_s = -234.0$, and $\eta_r = 0$.  In the figures on the right, the values of the constants are: $\eta_{\Lambda} = 1/3$, $\eta_c = 1$, $\eta_s = +234.0$, and $\eta_r = 0$. Values of parameters  taken from~\citep{Cordero2019}.} \label{LeftRight3}
\end{figure*} 

\subsection{Complex conjugate wave equations}

In the branch-cut cosmology, the Friedmann-type  equations analytically continued to the complex plane are~\citep{Zen2020,Zen2021a,Zen2021b}:
\begin{equation}
\Biggl(\frac{\frac{d}{dt} \ln^{-1}[\beta(t)]}{\ln^{-1}[\beta(t)]  } \Biggr)^2   =    \frac{8 \pi G}{3} \rho(t) 
-  \frac{kc^2}{\ln^{-1}[\beta(t)]} + \frac{1}{3} \Lambda \, ;  \label{NFE1} 
\end{equation}
and 
\begin{equation}
\Biggl( \frac{\frac{d^2}{dt^2} \ln^{-1}[\beta(t)]}{\ln^{-1}[\beta(t)] } \Biggr)   =  - \frac{4 \pi G}{3} \Big(\rho(t) + \frac{3}{c^2} p(t) \Bigr)
+  \frac{1}{3} \Lambda 
 ,  \label{NFE2}
\end{equation}
where $\Lambda$ represents the cosmological constant. The corresponding complex conjugated Friedmann's-type  equations are:
\begin{equation}
 \Biggl(\frac{\frac{d}{dt} \ln^{-1}(\beta^*(t^*))}{\ln^{-1}(\beta^*(t^*))} \Biggr)^2   =     \frac{8 \pi G}{3} \rho^*(t^*)
-  \frac{kc^2}{\ln^{-2}(\beta^*(t^*))} + \frac{1}{3} \Lambda^* ,   \label{NCFE1} 
\end{equation}
and
\begin{equation}
  \Biggl( \frac{\frac{d^2}{dt^{*2}} \ln^{-1}(\beta^*(t^*))}{\ln^{-1}(\beta^*(t^*)) } \Biggr)    =   - \frac{4 \pi G}{3} \Biggl(  \rho^*(t^*)  +  \frac{3}{c^2} p^*(t^*)  \Biggr)
+  \frac{1}{3} \Lambda^*  .   \label{NCFE2}
\end{equation}

Following a similar procedure, in this work we consider the complex conjugate version of equation (\ref{WDWS})

\begin{equation}
 \Biggl(- \frac{d^2}{d u^{*2}} +  \eta_c u^{*2}  - \eta_{\Lambda} u^{*4} 
 - \eta_r   -  \frac{\eta_s}{u^{*2}}  \Biggr) \Psi^*(u^*) = 0 \, . \label{WDWScc}
 \end{equation}

\subsection{Solutions and Boundary Conditions}

The most known proposals for the boundary conditions of the wave-function of the universe are the “no boundary”~\citep{Hartle} and tunneling boundary conditions~\citep{Vilenkin}. These boundary conditions, although based on different conceptions and assumptions, have in common the prediction of an inflationary stage of evolution of the Universe.  According to the singularity theorems of Hawking and Penrose, the universe would not have a classical beginning corresponding to a Lorentzian geometry, but would have started based on a regular Euclidean geometry with four spatial dimensions, followed by a quantum transition to a Lorentzian geometry with a temporal and three spatial dimensions~\citep{Hawking1982}. The tunneling boundary condition of \citet{Vilenkin1986} has two degrees of freedom: the scale factor and a homogeneous scalar field. A tunneling wave function then describes an ensemble of universes tunneling from ‘‘nothing’’ to a de Sitter space, and then evolving along the lines of an inflationary scenario and eventually collapsing to a singularity~\citep{Vilenkin1986}.

In this contribution, we present a different proposal, as we will see below, in order to reconcile the evolutionary behavior of the branch-cut universe in the anterior and posterior regions dominated by Planck dimensions. 

In Figures \ref{Sol1} and \ref{Sol2} we show a family of solutions of Equation \ref{WDWS}, considering a real potential, for the positive region of evolution of the branch-cut universe. The values of the constants are: $\eta_{\Lambda} = 1/3$, $\eta_c = 1$, $\eta_s = +0.03$, and $\eta_r = 0.6$. The boundary conditions correspond to $y(1) = 1$ and $y(1) = 0$, as for values of $x$ between 0 and 1 the equation has no solution. In the case where $x$ is negative, the solutions are similar but inverted, with a region between $x= -1$ and $x=1$ where the system of equations has no solution, which can be interpreted as the region where a quantum leap of the universe's wave function occurs. The numerical values correspond to natural units, and the factor $1$ can be interpreted as a Planck unit of length, or, in view of the Bekenstein criterion, the smallest unit of length corresponding to the region in which the criterion is satisfied.
In Figures \ref{LeftRight}, \ref{LeftRight2}, and \ref{LeftRight3}, we show typical solutions of Equations \ref{WDWS} and \ref{WDWScc}. The  boundary conditions are $\Psi(-1)=-1$ or 
$\Psi(-1)= 0$
on the negative sector, depending of the degree of convergence of the equations  and $\Psi(1) = 1$ or $\Psi(1)=0$ on the positive sector with the parameters taken from~\citep{Cordero2019}. As shown in the figures, for the region domains between $u = -1$ and $u=1$ the equations \ref{WDWS} and \ref{WDWScc}  have no solution. In our interpretation, this domain corresponds to the region in which a topological quantum leap occurs in accordance with the Bekenstein criterion~\citep{Zen2023a,Zen2023b}.

At this point, a  word of caution regarding the choice of parameters. In this work we seek to adopt more conventional values of parameters, standards we could say, focusing more on the structural aspects of the branch-cut cosmology solutions in order to facilitate comparisons with more conventional approaches. In the future, we intend to focus on the issue of parametric choices, adopting specific procedures, more in line with the concepts proposed by the model. In this research work we adopted the choices made by~\citet{Cordero2019}. For their choices, the authors considered different scenarios for the universe and for the potentials in the Wheeler-DeWitt equation: with an embryonic epoch, with a potential barrier and no embryonic epoch, without a potential barrier and a big bounce, with an initial singularity of the primordial universe and no potential barrier (for more details see~\citep{Cordero2019}).

\section{Final remarks}

Given the Klein-Gordon equation for a particle with mass $m$
\begin{equation}
\Bigl( \partial_{\mu} \partial^{\mu} + m^2 \Bigr) \Phi(x) = 0 \, , \label{DE}
\end{equation}
in covariant notation, the four-current (\ref{J}) of the Klein-Gordon equation may be written as
\begin{equation}
j^{\mu}(x) = i \Bigl[\Phi^*(x) \partial^{\mu} \Phi(x) - \Bigl( \partial^{\mu} \Phi^*(x)\Bigr)  \Phi(x)   \Bigr] . \label{Jc}
\end{equation}
Combining (\ref{DE}) and (\ref{Jc}), $\partial_ {\mu} j^{\mu}(x) =0$, so  
the four-current $j^{\mu}(x)$ is a conserved quantity. However, the density
\begin{equation}
\rho(x) = i \Bigl[\Phi^*(x) \frac{\partial \Phi(x)}{\partial t}  - \Bigl(\frac{\partial \Phi^*(x)}{\partial t} \Bigr)  \Phi(x)   \Bigr] , \label{rhoc}
\end{equation}
and the corresponding inner product
\begin{equation}
\langle \Phi_1(x) | \Phi_2(x)\rangle = \frac{i}{2} \int_{{\cal R}^3} d^3x \Bigl[\Phi_2(x) \frac{\partial \Phi_1(x)}{\partial t}  - \Bigl(\frac{\partial \Phi_2(x)}{\partial t} \Bigr)  \Phi_1(x)   \Bigr]\,, \label{inner}
\end{equation}
are not positive defined, so the Klein-Gordon wave function cannot therefore be interpreted as a probability amplitude. Moreover, endowing the vector space ${\cal V}$ of solutions of the Klein-Gordon equation with the ``inner product'' given by equation (\ref{inner}) does not generate a genuine inner product space. 
A similar problem arises in handling the Wheeler-DeWitt equation due to its Klein-Gordon structure, generating divergent interpretations about a probabilistic interpretation of the wave function of the universe. 

In this work, we do not intend to focus on the nature of these interpretations, but rather focus on some still open questions that may shed some light on the contribution about this topic the branch-cut cosmology can offer. For example, going back to the Wheeler-DeWitt equation and the non-definite positiveness of its inner product, a way out of this problem would be to restrict the domain of the inner product to a subspace of quantum state vectors that correspond to positive energy solutions. This procedure makes it possible to obtain a definite positive inner product that can be extended to a separable Hilbert space by means of Cauchy completion (see \citet{Mostafazadeh} and references therein). 

Another way of examining the same problem corresponds to the quantum field theory domain, in which solutions of the Klein-Gordon equation are quantized and turned out to field operators, supplying the formulation with a second quantization methodology. In this case, while the branch-cut equation can describe the evolution of quantum states ``with negative probability of realization'', the quantum field theory procedure describes the evolution of field operators associated to real and underlying virtual universes.  

More specifically, instead of seeking the realization of an explicit manifestly covariant description for a positive-definite and Lorentz-invariant inner product on the space of solutions of the Wheeler-DeWitt equation, 
with may adopt a description based on a kind of `third' quantization. In this procedure, 
we may complement the branch-cut wave function, solution of a formulation in the style of the Wheeler-DeWitt equation, with an operator structure of creation and annihilation operators of many universes. In short, while in quantum field theory, solutions to the Klein-Gordon equation are quantized and transformed into field operators, following a similar procedure, the $\Psi(u)$ solution of the branch-cut equation developed in the style of the Wheeler-DeWitt formulation, could be quantized and transformed into an operator that creates and annihilates universes, similarly to conventional particle creation and annihilation operators. The challenge of this proposition is to determine observables that quantify and justify this approach.

Hawking's interpretative proposal~\citep{Hartle}, in turn, assumes that $|\Psi[h_{ij}, \Phi,\Sigma)|^2$ is proportional to the probability $P_{\cal V}$ to find the universe in a constrained to a region configuration
${\cal V}$ of a superspace that contains a three-surface $\Sigma$, where the metric is described by $h_{ij}$ and matter fields represented by $\Phi$:
\begin{equation}
P_{\cal V} \propto \int_{{\cal V}} \Psi^*[h_{ij}, \Phi,\Sigma)] \Psi[h_{ij}, \Phi,\Sigma)] d{ \cal V} \, . 
\end{equation}

An attractive possibility would be to seek an interpretation that presents points of similarity with the view of \citet{Hartle} for the wave function of the universe, adapting it to those aspects that are inherent to branch-cut cosmology. 

In this interpretative search, however, we found also close contact points of the branch-cut cosmology with the Many-Worlds Interpretation of Quantum Mechanics, going back to Hugh~\citet{Everett}. According to him, there would be myriads of worlds in the Universe beyond the world we know. And when a quantum experiment with different possible outcomes is performed, all the outcomes are then realized, each of which is materialized in a newly created different world, even if we are only aware of the outcomes obtained in the world we live in\footnote{There is an aspect present in this formulation that refers to Plato's cosmological philosophical view that distinguished between appearance and reality while describing the systems of the universe~\citep{Plato}. Everett went further in his vision of quantum mechanics when he boldly invested in the vision of a wave mechanics without probability~\citep{Everett1955,Everett1957}.}.  
 
 The classical solution of the branch-cut cosmology, given by the complex form factor  
$\ln[\beta(t)]$, which  corresponds to a helix-like superposition of cut-planes, maps an infinite number of  Riemann sheets onto horizontal strips, which represent the complex time evolution of horizon sizes. The patch sizes in turn maps progressively the various branches of the $\ln[\beta(t)]$ function which are {\it glued} along the copies of each upper-half plane with their copies on the corresponding lower-half planes. In the branch-cut cosmology, the cosmic singularity is replaced by a family of Riemann sheets in which the scale factor shrinks to a finite critical size, --- the range of $\ln^{-1}[\beta(t)]$, associated to the cuts in the branch cut, shaped by the $\beta(t)$ function ---, well above the Planck length. 
In the contraction phase, 
as the patch size decreases with a linear dependence on $\ln[\beta(t)]$, light travels through geodesics on each Riemann sheet,  circumventing continuously the branch-cut,  and although the horizon size scale with $\ln^{\epsilon}[\beta(t)]$, the length of the path to be traveled by light compensates for the scaling difference between the patch and horizon sizes. Under these conditions, causality between the horizon size and the patch size may be achieved through the accumulation of branches in the transition region between the present state of the universe and the past events. In the branch-cut cosmology conception, the cosmological arrow of time is determined as the direction in which ``time'', from the macroscopic point of view, flows globally. In the branch-cut cosmology, in turn, this local component is identified with the corresponding real local time component that results from the complexification of the FLRW metric ~\citep{Zen2023b}.

Following previous proposals, $\Psi(\ln[\beta(t)])$ may be interpreted as the probability amplitude and its modulus square as the probability density distribution of universes for a given family of branches characterized by  the complex form factor $\ln[\beta(t)]$. The wave function shows oscillations, i.e., its absolute value has maxima, so the corresponding values of $\ln[\beta(t)]$ are the most probable universes constricted to the branches 
$|\ln[\beta(t)]|$. Thus, there is a quasi-discrete distribution of universes corresponding to different value of the complex branch-cut form factor. The results also indicate that the wave function is suppressed toward small $|\ln[\beta(t)]|$ which may imply that very small universes are suppressed.

Many relevant questions involving the quantum version of branch-cut cosmology are still open, such as the role of complex potentials, the presence of higher spatial curvature terms in the formulation, the quantization of helix-type solutions, more in tune with the classical formulation, a more consistent evolution mapping of the cut-planes, the consequences of adopting a non-symmetric approach and a different ordering of the dimensionless thermodynamics connection $\epsilon$, the role of dark matter in the
evolution of the branch-cut universe, the role of fluctuations in the primordial spectrum and seeds in the 
the early
universe, and also questions regarding the multiverse
content, as well as a class of solutions with ordering parameters (see \citet{Vieira}) among others.

\section{Acknowledgements}
P.O.H. acknowledges financial support from PAPIIT-DGAPA (IN100421).  The authors wish to thank the referees for valuable comments.

\bibliography{ZenB}%

\begin{thebibliography}{}

\bibitem [\protect \citeauthoryear {%
Arnowitt%
, Deser%
\BCBL {}\ \BBA {} Misner%
}{%
Arnowitt%
\ \protect \BOthers {.}}{%
{\protect \APACyear {2004}}%
}]{%
ADM}
\APACinsertmetastar {%
ADM}%
\begin{APACrefauthors}%
Arnowitt, R.%
, Deser, S.%
\BCBL {}\ \BBA {} Misner, C.%
\end{APACrefauthors}%
\unskip\
\newblock
\APACrefYearMonthDay{2004}{}{},
\newblock
\APACrefbtitle {The Dynamics of General Relativity.} {The Dynamics of General
  Relativity.}
\newblock
\APACrefnote{arXiv:gr-qc/0405109}
\PrintBackRefs{\CurrentBib}

\bibitem [\protect \citeauthoryear {%
Bertolami%
\ \BBA {} Zarro%
}{%
Bertolami%
\ \BBA {} Zarro%
}{%
{\protect \APACyear {2011}}%
}]{%
Bertolami2011}
\APACinsertmetastar {%
Bertolami2011}%
\begin{APACrefauthors}%
Bertolami, O.%
\BCBT {}\ \BBA {} Zarro, C\BPBI A.%
\end{APACrefauthors}%
\unskip\
\newblock
\APACrefYearMonthDay{2011}{}{},
\newblock
\unskip
\newblock
\APACjournalVolNumPages{Phys. Rev. D}{84}{}{044042}.
\PrintBackRefs{\CurrentBib}

\bibitem [\protect \citeauthoryear {%
Bodmann%
, Zen~Vasconcellos%
, de Freitas~Pacheco%
, Hess%
\BCBL {}\ \BBA {} Hadjimichef%
}{%
Bodmann%
\ \protect \BOthers {.}}{%
{\protect \APACyear {2023}}%
}]{%
Zen2023b}
\APACinsertmetastar {%
Zen2023b}%
\begin{APACrefauthors}%
Bodmann, B.%
, Zen~Vasconcellos, C\BPBI A.%
, de Freitas~Pacheco, J\BPBI A.%
, Hess, P\BPBI O.%
\BCBL {}\ \BBA {} Hadjimichef, D.%
\end{APACrefauthors}%
\unskip\
\newblock
\APACrefYearMonthDay{2023}{}{},
\newblock
\unskip
\newblock
\APACjournalVolNumPages{Astronomische Nachrichten}{}{}{}.
\newblock
\APACrefnote{To be published.}
\PrintBackRefs{\CurrentBib}

\bibitem [\protect \citeauthoryear {%
Chojnacki%
\ \BBA {} Kwapisz%
}{%
Chojnacki%
\ \BBA {} Kwapisz%
}{%
{\protect \APACyear {2021}}%
}]{%
Chojnacki}
\APACinsertmetastar {%
Chojnacki}%
\begin{APACrefauthors}%
Chojnacki, J.%
\BCBT {}\ \BBA {} Kwapisz, J.%
\end{APACrefauthors}%
\unskip\
\newblock
\APACrefYearMonthDay{2021}{}{},
\newblock
\unskip
\newblock
\APACjournalVolNumPages{Phys.Rev.D}{104}{}{103504}.
\PrintBackRefs{\CurrentBib}

\bibitem [\protect \citeauthoryear {%
Cordero%
, Garcia-Compean%
\BCBL {}\ \BBA {} Turrubiates%
}{%
Cordero%
\ \protect \BOthers {.}}{%
{\protect \APACyear {2019}}%
}]{%
Cordero2019}
\APACinsertmetastar {%
Cordero2019}%
\begin{APACrefauthors}%
Cordero, R.%
, Garcia-Compean, H.%
\BCBL {}\ \BBA {} Turrubiates, F\BPBI J.%
\end{APACrefauthors}%
\unskip\
\newblock
\APACrefYearMonthDay{2019}{}{},
\newblock
\unskip
\newblock
\APACjournalVolNumPages{General Relativity and Gravitation}{51}{}{138}.
\PrintBackRefs{\CurrentBib}

\bibitem [\protect \citeauthoryear {%
de Freitas~Pacheco%
, Zen~Vasconcellos%
, Hadjimichef%
, Hess%
\BCBL {}\ \BBA {} Bodmann%
}{%
de Freitas~Pacheco%
\ \protect \BOthers {.}}{%
{\protect \APACyear {2023}}%
}]{%
Zen2023a}
\APACinsertmetastar {%
Zen2023a}%
\begin{APACrefauthors}%
de Freitas~Pacheco, J\BPBI A.%
, Zen~Vasconcellos, C\BPBI A.%
, Hadjimichef, D.%
, Hess, P\BPBI O.%
\BCBL {}\ \BBA {} Bodmann, B.%
\end{APACrefauthors}%
\unskip\
\newblock
\APACrefYearMonthDay{2023}{}{},
\newblock
\APACrefbtitle {Branch-Cut Cosmology and the Bekenstein Criterion.} {Branch-Cut
  Cosmology and the Bekenstein Criterion.}
\newblock
\APACrefnote{To be published by the Astronomische Nachrichten}
\PrintBackRefs{\CurrentBib}

\bibitem [\protect \citeauthoryear {%
DeWitt%
}{%
DeWitt%
}{%
{\protect \APACyear {1967}}%
}]{%
WDW}
\APACinsertmetastar {%
WDW}%
\begin{APACrefauthors}%
DeWitt, B\BPBI S.%
\end{APACrefauthors}%
\unskip\
\newblock
\APACrefYearMonthDay{1967}{}{},
\newblock
\unskip
\newblock
\APACjournalVolNumPages{Phys. Rev.}{160}{}{1113}.
\PrintBackRefs{\CurrentBib}

\bibitem [\protect \citeauthoryear {%
Everett~III%
}{%
Everett~III%
}{%
{\protect \APACyear {1956}}%
}]{%
Everett}
\APACinsertmetastar {%
Everett}%
\begin{APACrefauthors}%
Everett~III, H.%
\end{APACrefauthors}%
\unskip\
\newblock
\APACrefYear{1956}.
\unskip\
\newblock
\APACrefbtitle {The Many-Worlds Interpretation of Quantum Mechanics} {The
  Many-Worlds Interpretation of Quantum Mechanics}\ \APACtypeAddressSchool
  {\BUPhD}{}{}.
\unskip\
\newblock
\APACaddressSchool {Princeton, USA}{Princeton University Press}.
\PrintBackRefs{\CurrentBib}

\bibitem [\protect \citeauthoryear {%
Everett~III%
}{%
Everett~III%
}{%
{\protect \APACyear {1957}}%
}]{%
Everett1957}
\APACinsertmetastar {%
Everett1957}%
\begin{APACrefauthors}%
Everett~III, H.%
\end{APACrefauthors}%
\unskip\
\newblock
\APACrefYearMonthDay{1957}{}{},
\newblock
\unskip
\newblock
\APACjournalVolNumPages{Reviews of Modern Physics}{29 (3)}{}{454-462}.
\PrintBackRefs{\CurrentBib}

\bibitem [\protect \citeauthoryear {%
Everett~III%
}{%
Everett~III%
}{%
{\protect \APACyear {1975 [1955]}}%
}]{%
Everett1955}
\APACinsertmetastar {%
Everett1955}%
\begin{APACrefauthors}%
Everett~III, H.%
\end{APACrefauthors}%
\unskip\
\newblock
\APACrefYear{1975 [1955]},
\newblock
\APACrefbtitle {The Theory of the Universal Wave Function} {The Theory of the
  Universal Wave Function}.
\PrintBackRefs{\CurrentBib}

\bibitem [\protect \citeauthoryear {%
Feinberg%
\ \BBA {} Peleg%
}{%
Feinberg%
\ \BBA {} Peleg%
}{%
{\protect \APACyear {1995}}%
}]{%
Feinberg}
\APACinsertmetastar {%
Feinberg}%
\begin{APACrefauthors}%
Feinberg, J.%
\BCBT {}\ \BBA {} Peleg, Y.%
\end{APACrefauthors}%
\unskip\
\newblock
\APACrefYearMonthDay{1995}{}{},
\newblock
\unskip
\newblock
\APACjournalVolNumPages{Phys.Rev. D}{52}{}{1988}.
\PrintBackRefs{\CurrentBib}

\bibitem [\protect \citeauthoryear {%
Flinn%
}{%
Flinn%
}{%
{\protect \APACyear {2020}}%
}]{%
Flinn}
\APACinsertmetastar {%
Flinn}%
\begin{APACrefauthors}%
Flinn, G.%
\end{APACrefauthors}%
\unskip\
\newblock
\APACrefYear{2020}.
\unskip\
\newblock
\APACrefbtitle {The Wave Function of The Universe} {The Wave Function of The
  Universe}\ \APACtypeAddressSchool {\BUPhD}{}{}.
\unskip\
\newblock
\APACaddressSchool {London, UK}{Imperial College}.
\PrintBackRefs{\CurrentBib}

\bibitem [\protect \citeauthoryear {%
Garattini%
\ \BBA {} Faizal%
}{%
Garattini%
\ \BBA {} Faizal%
}{%
{\protect \APACyear {2016}}%
}]{%
Garattini2016}
\APACinsertmetastar {%
Garattini2016}%
\begin{APACrefauthors}%
Garattini, R.%
\BCBT {}\ \BBA {} Faizal, M.%
\end{APACrefauthors}%
\unskip\
\newblock
\APACrefYearMonthDay{2016}{}{},
\newblock
\unskip
\newblock
\APACjournalVolNumPages{Nuclear Physics B}{905}{}{313-326}.
\PrintBackRefs{\CurrentBib}

\bibitem [\protect \citeauthoryear {%
Garc\'ia-Compe\'an%
\ \BBA {} Mata-Pacheco%
}{%
Garc\'ia-Compe\'an%
\ \BBA {} Mata-Pacheco%
}{%
{\protect \APACyear {2022}}%
}]{%
Compean}
\APACinsertmetastar {%
Compean}%
\begin{APACrefauthors}%
Garc\'ia-Compe\'an, H.%
\BCBT {}\ \BBA {} Mata-Pacheco, D.%
\end{APACrefauthors}%
\unskip\
\newblock
\APACrefYearMonthDay{2022}{}{},
\newblock
\unskip
\newblock
\APACjournalVolNumPages{Universe}{8}{}{237}.
\PrintBackRefs{\CurrentBib}

\bibitem [\protect \citeauthoryear {%
Hartle%
\ \BBA {} Hawking%
}{%
Hartle%
\ \BBA {} Hawking%
}{%
{\protect \APACyear {1983}}%
}]{%
Hartle}
\APACinsertmetastar {%
Hartle}%
\begin{APACrefauthors}%
Hartle, J.%
\BCBT {}\ \BBA {} Hawking, S\BPBI W.%
\end{APACrefauthors}%
\unskip\
\newblock
\APACrefYearMonthDay{1983}{}{},
\newblock
\unskip
\newblock
\APACjournalVolNumPages{Phys. Rev. D}{28}{}{2960}.
\PrintBackRefs{\CurrentBib}

\bibitem [\protect \citeauthoryear {%
Hawking%
}{%
Hawking%
}{%
{\protect \APACyear {1982}}%
}]{%
Hawking1982}
\APACinsertmetastar {%
Hawking1982}%
\begin{APACrefauthors}%
Hawking, S.%
\end{APACrefauthors}%
\unskip\
\newblock
\APACrefYearMonthDay{1982}{}{},
\newblock
\unskip
\newblock
\APACjournalVolNumPages{Pontif. Acad. Sci. Scr. Varia}{48}{}{563}.
\PrintBackRefs{\CurrentBib}

\bibitem [\protect \citeauthoryear {%
Ho\v{r}ava%
}{%
Ho\v{r}ava%
}{%
{\protect \APACyear {2009}}%
}]{%
Horava}
\APACinsertmetastar {%
Horava}%
\begin{APACrefauthors}%
Ho\v{r}ava, P.%
\end{APACrefauthors}%
\unskip\
\newblock
\APACrefYearMonthDay{2009}{}{},
\newblock
\unskip
\newblock
\APACjournalVolNumPages{Phys. Rev. D}{79}{}{084008}.
\PrintBackRefs{\CurrentBib}

\bibitem [\protect \citeauthoryear {%
i. Maeda%
, Misonoh%
\BCBL {}\ \BBA {} Kobayashi%
}{%
i. Maeda%
\ \protect \BOthers {.}}{%
{\protect \APACyear {2010}}%
}]{%
Maeda}
\APACinsertmetastar {%
Maeda}%
\begin{APACrefauthors}%
i. Maeda, K.%
, Misonoh, Y.%
\BCBL {}\ \BBA {} Kobayashi, T.%
\end{APACrefauthors}%
\unskip\
\newblock
\APACrefYearMonthDay{2010}{}{},
\newblock
\unskip
\newblock
\APACjournalVolNumPages{Phys. Rev. D}{82}{}{}.
\PrintBackRefs{\CurrentBib}

\bibitem [\protect \citeauthoryear {%
Kim%
}{%
Kim%
}{%
{\protect \APACyear {1997}}%
}]{%
Kim}
\APACinsertmetastar {%
Kim}%
\begin{APACrefauthors}%
Kim, S\BPBI P.%
\end{APACrefauthors}%
\unskip\
\newblock
\APACrefYearMonthDay{1997}{}{},
\newblock
\unskip
\newblock
\APACjournalVolNumPages{Phys. Lett. A}{236}{}{11}.
\PrintBackRefs{\CurrentBib}

\bibitem [\protect \citeauthoryear {%
Mostafazadeh%
\ \BBA {} Zamani%
}{%
Mostafazadeh%
\ \BBA {} Zamani%
}{%
{\protect \APACyear {2006}}%
}]{%
Mostafazadeh}
\APACinsertmetastar {%
Mostafazadeh}%
\begin{APACrefauthors}%
Mostafazadeh, A.%
\BCBT {}\ \BBA {} Zamani, F.%
\end{APACrefauthors}%
\unskip\
\newblock
\APACrefYearMonthDay{2006}{}{},
\newblock
\unskip
\newblock
\APACjournalVolNumPages{Annals of Physics}{321 (9)}{}{2183-2209}.
\PrintBackRefs{\CurrentBib}

\bibitem [\protect \citeauthoryear {%
Mukohyama%
}{%
Mukohyama%
}{%
{\protect \APACyear {2009}}%
}]{%
Mukohyama2009}
\APACinsertmetastar {%
Mukohyama2009}%
\begin{APACrefauthors}%
Mukohyama, S.%
\end{APACrefauthors}%
\unskip\
\newblock
\APACrefYearMonthDay{2009}{}{},
\newblock
\unskip
\newblock
\APACjournalVolNumPages{Phys. Rev. D}{80}{}{064005}.
\PrintBackRefs{\CurrentBib}

\bibitem [\protect \citeauthoryear {%
Mukohyama%
}{%
Mukohyama%
}{%
{\protect \APACyear {2010}}%
}]{%
Mukohyama}
\APACinsertmetastar {%
Mukohyama}%
\begin{APACrefauthors}%
Mukohyama, S.%
\end{APACrefauthors}%
\unskip\
\newblock
\APACrefYearMonthDay{2010}{}{},
\newblock
\unskip
\newblock
\APACjournalVolNumPages{Class. Quant. Grav.}{27}{}{223101}.
\PrintBackRefs{\CurrentBib}

\bibitem [\protect \citeauthoryear {%
Plato%
}{%
Plato%
}{%
{\protect \APACyear {[428 - 348 a.C.] 2004}}%
}]{%
Plato}
\APACinsertmetastar {%
Plato}%
\begin{APACrefauthors}%
Plato.%
\end{APACrefauthors}%
\unskip\
\newblock
\APACrefYearMonthDay{[428 - 348 a.C.] 2004}{}{},
\newblock
\APACrefbtitle {Plato - Stanford Encyclopedia of Philosophy.} {Plato - Stanford
  Encyclopedia of Philosophy.}
\newblock
\APACrefnote{\url{https://plato.stanford.edu/entries/plato/}}
\PrintBackRefs{\CurrentBib}

\bibitem [\protect \citeauthoryear {%
Rovelli%
}{%
Rovelli%
}{%
{\protect \APACyear {2015}}%
}]{%
Rovelli2015}
\APACinsertmetastar {%
Rovelli2015}%
\begin{APACrefauthors}%
Rovelli, C.%
\end{APACrefauthors}%
\unskip\
\newblock
\APACrefYearMonthDay{2015}{}{},
\newblock
\unskip
\newblock
\APACjournalVolNumPages{Classical and Quantum Gravity}{32 (12)}{}{124005}.
\PrintBackRefs{\CurrentBib}

\bibitem [\protect \citeauthoryear {%
Rovelli%
}{%
Rovelli%
}{%
{\protect \APACyear {2019}}%
}]{%
Rovelli2019}
\APACinsertmetastar {%
Rovelli2019}%
\begin{APACrefauthors}%
Rovelli, C.%
\end{APACrefauthors}%
\unskip\
\newblock
\APACrefYear{2019},
\newblock
\APACrefbtitle {The Order of Time} {The Order of Time}.
\newblock
\APACaddressPublisher{New York, USA}{Riverhead Books}.
\PrintBackRefs{\CurrentBib}

\bibitem [\protect \citeauthoryear {%
Rovelli%
\ \BBA {} Smerlak%
}{%
Rovelli%
\ \BBA {} Smerlak%
}{%
{\protect \APACyear {2011}}%
}]{%
Rovelli2011}
\APACinsertmetastar {%
Rovelli2011}%
\begin{APACrefauthors}%
Rovelli, C.%
\BCBT {}\ \BBA {} Smerlak, M.%
\end{APACrefauthors}%
\unskip\
\newblock
\APACrefYearMonthDay{2011}{}{},
\newblock
\unskip
\newblock
\APACjournalVolNumPages{Classical and Quantum Gravity}{28}{}{075007}.
\PrintBackRefs{\CurrentBib}

\bibitem [\protect \citeauthoryear {%
Saridakis%
}{%
Saridakis%
}{%
{\protect \APACyear {2010}}%
}]{%
Saridakis}
\APACinsertmetastar {%
Saridakis}%
\begin{APACrefauthors}%
Saridakis, E\BPBI N.%
\end{APACrefauthors}%
\unskip\
\newblock
\APACrefYearMonthDay{2010}{}{},
\newblock
\APACrefbtitle {Aspects of Horava-Lifshitz cosmology.} {Aspects of
  Horava-Lifshitz cosmology.}
\newblock
\APACrefnote{arXiv:1101.0300 [astro-ph.CO]}
\PrintBackRefs{\CurrentBib}

\bibitem [\protect \citeauthoryear {%
T.~Shestakova%
}{%
T.~Shestakova%
}{%
{\protect \APACyear {2019}}%
}]{%
Shestakova2019}
\APACinsertmetastar {%
Shestakova2019}%
\begin{APACrefauthors}%
Shestakova, T.%
\end{APACrefauthors}%
\unskip\
\newblock
\APACrefYearMonthDay{2019}{}{},
\newblock
\unskip
\newblock
\APACjournalVolNumPages{Int. J. Mod. Phys. D}{28 (13)}{}{1941009}.
\PrintBackRefs{\CurrentBib}

\bibitem [\protect \citeauthoryear {%
T\BPBI P.~Shestakova%
}{%
T\BPBI P.~Shestakova%
}{%
{\protect \APACyear {2018}}%
}]{%
Shestakova2018}
\APACinsertmetastar {%
Shestakova2018}%
\begin{APACrefauthors}%
Shestakova, T\BPBI P.%
\end{APACrefauthors}%
\unskip\
\newblock
\APACrefYearMonthDay{2018}{}{},
\newblock
\unskip
\newblock
\APACjournalVolNumPages{Int. J. Mod. Phys. D}{27}{}{1841004}.
\PrintBackRefs{\CurrentBib}

\bibitem [\protect \citeauthoryear {%
Vakili%
\ \BBA {} Kord%
}{%
Vakili%
\ \BBA {} Kord%
}{%
{\protect \APACyear {2013}}%
}]{%
Vakili}
\APACinsertmetastar {%
Vakili}%
\begin{APACrefauthors}%
Vakili, B.%
\BCBT {}\ \BBA {} Kord, V.%
\end{APACrefauthors}%
\unskip\
\newblock
\APACrefYearMonthDay{2013}{}{},
\newblock
\unskip
\newblock
\APACjournalVolNumPages{Gen. Rel. Grav.}{45}{}{1313}.
\PrintBackRefs{\CurrentBib}

\bibitem [\protect \citeauthoryear {%
Vieira%
, Bezerra%
, Muniz%
, Cunha%
\BCBL {}\ \BBA {} Christiansen%
}{%
Vieira%
\ \protect \BOthers {.}}{%
{\protect \APACyear {2020}}%
}]{%
Vieira}
\APACinsertmetastar {%
Vieira}%
\begin{APACrefauthors}%
Vieira, H.%
, Bezerra, V.%
, Muniz, C.%
, Cunha, M.%
\BCBL {}\ \BBA {} Christiansen, H.%
\end{APACrefauthors}%
\unskip\
\newblock
\APACrefYearMonthDay{2020}{}{},
\newblock
\unskip
\newblock
\APACjournalVolNumPages{Phys. Lett. B}{809}{}{135712}.
\PrintBackRefs{\CurrentBib}

\bibitem [\protect \citeauthoryear {%
Vilenkin%
}{%
Vilenkin%
}{%
{\protect \APACyear {1982}}%
}]{%
Vilenkin}
\APACinsertmetastar {%
Vilenkin}%
\begin{APACrefauthors}%
Vilenkin, A.%
\end{APACrefauthors}%
\unskip\
\newblock
\APACrefYearMonthDay{1982}{}{},
\newblock
\unskip
\newblock
\APACjournalVolNumPages{Phys. Lett. B}{117}{}{25}.
\PrintBackRefs{\CurrentBib}

\bibitem [\protect \citeauthoryear {%
Vilenkin%
}{%
Vilenkin%
}{%
{\protect \APACyear {1986}}%
}]{%
Vilenkin1986}
\APACinsertmetastar {%
Vilenkin1986}%
\begin{APACrefauthors}%
Vilenkin, A.%
\end{APACrefauthors}%
\unskip\
\newblock
\APACrefYearMonthDay{1986}{}{},
\newblock
\unskip
\newblock
\APACjournalVolNumPages{Phys. Rev. D}{33}{}{3560}.
\PrintBackRefs{\CurrentBib}

\bibitem [\protect \citeauthoryear {%
Wiltshire%
}{%
Wiltshire%
}{%
{\protect \APACyear {1996}}%
}]{%
Wiltshire}
\APACinsertmetastar {%
Wiltshire}%
\begin{APACrefauthors}%
Wiltshire, D.%
\end{APACrefauthors}%
\unskip\
\newblock
\APACrefYear{1996},
\newblock
\APACrefbtitle {An introduction to quantum cosmology} {An introduction to
  quantum cosmology}\ (\BVOL~2; B.~Robson, N.~Visvanathan\BCBL {}\ \BBA {}
  W.~Woolcock, \BEDS{}).
\newblock
\APACaddressPublisher{Singapore}{World Scientific Pub. Co.}
\PrintBackRefs{\CurrentBib}

\bibitem [\protect \citeauthoryear {%
Zen~Vasconcellos%
, Hadjimichef%
, Hess%
, de Freitas~Pacheco%
\BCBL {}\ \BBA {} Bodmann%
}{%
Zen~Vasconcellos%
\ \protect \BOthers {.}}{%
{\protect \APACyear {2022}}%
}]{%
Zen2022}
\APACinsertmetastar {%
Zen2022}%
\begin{APACrefauthors}%
Zen~Vasconcellos, C\BPBI A.%
, Hadjimichef, D.%
, Hess, P\BPBI O.%
, de Freitas~Pacheco, J\BPBI A.%
\BCBL {}\ \BBA {} Bodmann, B.%
\end{APACrefauthors}%
\unskip\
\newblock
\APACrefYearMonthDay{2022}{}{},
\newblock
\APACrefbtitle {Evidences for the Branch-Cut Cosmology.} {Evidences for the
  Branch-Cut Cosmology.}
\newblock
\APACrefnote{XXI Meeting of Physics. UNSAAC, Cusco, Per\'u, 16-18 December
  2021. To be published by Journal of Physics: Conference Series}
\PrintBackRefs{\CurrentBib}

\bibitem [\protect \citeauthoryear {%
Zen~Vasconcellos%
, Hadjimichef%
, Razeira%
, Volkmer%
\BCBL {}\ \BBA {} Bodmann%
}{%
Zen~Vasconcellos%
\ \protect \BOthers {.}}{%
{\protect \APACyear {2020}}%
}]{%
Zen2020}
\APACinsertmetastar {%
Zen2020}%
\begin{APACrefauthors}%
Zen~Vasconcellos, C\BPBI A.%
, Hadjimichef, D.%
, Razeira, M.%
, Volkmer, G.%
\BCBL {}\ \BBA {} Bodmann, B.%
\end{APACrefauthors}%
\unskip\
\newblock
\APACrefYearMonthDay{2020}{}{},
\newblock
\unskip
\newblock
\APACjournalVolNumPages{Astronomische Nachrichten}{340 (9,10)}{}{857}.
\PrintBackRefs{\CurrentBib}

\bibitem [\protect \citeauthoryear {%
Zen~Vasconcellos%
\ \protect \BOthers {.}}{%
Zen~Vasconcellos%
\ \protect \BOthers {.}}{%
{\protect \APACyear {2021}}%
{\protect \APACexlab {{\protect \BCnt {1}}}}}]{%
Zen2021b}
\APACinsertmetastar {%
Zen2021b}%
\begin{APACrefauthors}%
Zen~Vasconcellos, C\BPBI A.%
, Hess, P\BPBI O.%
, Hadjimichef, D.%
, Bodmann, B.%
, Razeira, M.%
\BCBL {}\ \BBA {} Volkmer, G.%
\end{APACrefauthors}%
\unskip\
\newblock
\APACrefYearMonthDay{2021{\protect \BCnt {1}}}{}{},
\newblock
\unskip
\newblock
\APACjournalVolNumPages{Astronomische Nachrichten}{342 (5)}{}{776-787}.
\PrintBackRefs{\CurrentBib}

\bibitem [\protect \citeauthoryear {%
Zen~Vasconcellos%
\ \protect \BOthers {.}}{%
Zen~Vasconcellos%
\ \protect \BOthers {.}}{%
{\protect \APACyear {2021}}%
{\protect \APACexlab {{\protect \BCnt {2}}}}}]{%
Zen2021a}
\APACinsertmetastar {%
Zen2021a}%
\begin{APACrefauthors}%
Zen~Vasconcellos, C\BPBI A.%
, Hess, P\BPBI O.%
, Hadjimichef, D.%
, Bodmann, B.%
, Razeira, M.%
\BCBL {}\ \BBA {} Volkmer, G.%
\end{APACrefauthors}%
\unskip\
\newblock
\APACrefYearMonthDay{2021{\protect \BCnt {2}}}{}{},
\newblock
\unskip
\newblock
\APACjournalVolNumPages{Astronomische Nachrichten}{342 (5)}{}{765-775}.
\PrintBackRefs{\CurrentBib}

\end{thebibliography}

\end{document}